\documentstyle[12pt]{article}

\catcode`\@=11
\long\def\@makefntext#1{ 
\protect\noindent \hbox to 3.2pt {\hskip-.9pt
$^{{\ninerm\@thefnmark}}$\hfil}#1\hfill} 

\def\thefootnote{\fnsymbol{footnote}}
 \def\@makefnmark{\hbox to 0pt{$^{\@thefnmark}$\hss}}  

\def\ps@myheadings{\let\@mkboth\@gobbletwo
\def\@oddhead{\hbox{} 
\rightmark\hfil\ninerm\thepage}
\def\@oddfoot{}\def\@evenhead{\ninerm\thepage\hfil 
\leftmark\hbox{}}\def\@evenfoot{}
\def\sectionmark##1{}\def\subsectionmark##1{}}

\begin{document}
\begin{flushright} PSU/TH/146\\
hep-th/9405335\\
May 1994  \end{flushright}

\newcommand{\symbolfootnote}{\renewcommand{\thefootnote}
	{\fnsymbol{footnote}}}
\renewcommand{\thefootnote}{\fnsymbol{footnote}}
\newcommand{\alphfootnote}
	{\setcounter{footnote}{0}
	 \renewcommand{\thefootnote}{\sevenrm\alph{footnote}}}
\newcommand{\lN}{\mbox{$\langle N|$}}   
\newcommand{\rN}{\mbox{$|N\rangle $}}   
\newcommand{\mael}{matrix element}
\newcommand{\crse}{cross section}
\newcommand{\II}{\mbox{$I\bar{I}$}}
\newcommand{\qcd}{quantum chromodynamics}   
\newcommand{\eea}{$e^{+}e^{-}$ annihilation}
\newcommand{\ipp}{inclusive particle production}
\newcommand{\dis}{deep inelastic scattering}
\newcommand{\ie}{\mbox{$i \epsilon $}}
\newcommand{\ieo}{\mbox{$i \epsilon x_{0}$}}
\newcommand{\lvac}{\mbox{$\langle 0|$}}   
\newcommand{\rvac}{\mbox{$|0\rangle $}}   

\newcounter{sectionc}\newcounter{subsectionc}\newcounter{subsubsectionc}
\renewcommand{\section}[1] {\vspace{0.6cm}\addtocounter{sectionc}{1}
\setcounter{subsectionc}{0}\setcounter{subsubsectionc}{0}\noindent
	{\bf\thesectionc. #1}\par\vspace{0.4cm}}
\renewcommand{\subsection}[1] {\vspace{0.6cm}\addtocounter{subsectionc}{1}
	\setcounter{subsubsectionc}{0}\noindent
	{\it\thesectionc.\thesubsectionc. #1}\par\vspace{0.4cm}}
\renewcommand{\subsubsection}[1]
{\vspace{0.6cm}\addtocounter{subsubsectionc}{1}
	\noindent {\rm\thesectionc.\thesubsectionc.\thesubsubsectionc.
	#1}\par\vspace{0.4cm}}
\newcommand{\nonumsection}[1] {\vspace{0.6cm}\noindent{\bf #1}
	\par\vspace{0.4cm}}

\newcounter{appendixc}
\newcounter{subappendixc}[appendixc]
\newcounter{subsubappendixc}[subappendixc]
\renewcommand{\thesubappendixc}{\Alph{appendixc}.\arabic{subappendixc}}
\renewcommand{\thesubsubappendixc}
	{\Alph{appendixc}.\arabic{subappendixc}.\arabic{subsubappendixc}}

\renewcommand{\appendix}[1] {\vspace{0.6cm}
        \refstepcounter{appendixc}
        \setcounter{figure}{0}
        \setcounter{table}{0}
        \setcounter{equation}{0}
        \renewcommand{\thefigure}{\Alph{appendixc}.\arabic{figure}}
        \renewcommand{\thetable}{\Alph{appendixc}.\arabic{table}}
        \renewcommand{\theappendixc}{\Alph{appendixc}}
        \renewcommand{\theequation}{\Alph{appendixc}.\arabic{equation}}
        \noindent{\bf Appendix \theappendixc #1}\par\vspace{0.4cm}}
\newcommand{\subappendix}[1] {\vspace{0.6cm}
        \refstepcounter{subappendixc}
        \noindent{\bf Appendix \thesubappendixc. #1}\par\vspace{0.4cm}}
\newcommand{\subsubappendix}[1] {\vspace{0.6cm}
        \refstepcounter{subsubappendixc}
        \noindent{\it Appendix \thesubsubappendixc. #1}
	\par\vspace{0.4cm}}

\def\abstracts#1{{
	\centering{\begin{minipage}{30pc}\tenrm\baselineskip=12pt\noindent
	\centerline{\tenrm ABSTRACT}\vspace{0.3cm}
	\parindent=0pt #1
	\end{minipage} }\par}}

\newcommand{\bibit}{\it}
\newcommand{\bibbf}{\bf}
\renewenvironment{thebibliography}[1]
	{\begin{list}{\arabic{enumi}.}
	{\usecounter{enumi}\setlength{\parsep}{0pt}
\setlength{\leftmargin 1.25cm}{\rightmargin 0pt}
	 \setlength{\itemsep}{0pt} \settowidth
	{\labelwidth}{#1.}\sloppy}}{\end{list}}

\topsep=0in\parsep=0in\itemsep=0in
\parindent=1.5pc

\newcounter{itemlistc}
\newcounter{romanlistc}
\newcounter{alphlistc}
\newcounter{arabiclistc}
\newenvironment{itemlist}
    	{\setcounter{itemlistc}{0}
	 \begin{list}{$\bullet$}
	{\usecounter{itemlistc}
	 \setlength{\parsep}{0pt}
	 \setlength{\itemsep}{0pt}}}{\end{list}}

\newenvironment{romanlist}
	{\setcounter{romanlistc}{0}
	 \begin{list}{$($\roman{romanlistc}$)$}
	{\usecounter{romanlistc}
	 \setlength{\parsep}{0pt}
	 \setlength{\itemsep}{0pt}}}{\end{list}}

\newenvironment{alphlist}
	{\setcounter{alphlistc}{0}
	 \begin{list}{$($\alph{alphlistc}$)$}
	{\usecounter{alphlistc}
	 \setlength{\parsep}{0pt}
	 \setlength{\itemsep}{0pt}}}{\end{list}}

\newenvironment{arabiclist}
	{\setcounter{arabiclistc}{0}
	 \begin{list}{\arabic{arabiclistc}}
	{\usecounter{arabiclistc}
	 \setlength{\parsep}{0pt}
	 \setlength{\itemsep}{0pt}}}{\end{list}}

\newcommand{\fcaption}[1]{
        \refstepcounter{figure}
        \setbox\@tempboxa = \hbox{\tenrm Fig.~\thefigure. #1}
        \ifdim \wd\@tempboxa > 6in
           {\begin{center}
        \parbox{6in}{\tenrm\baselineskip=12pt Fig.~\thefigure. #1 }
            \end{center}}
        \else
             {\begin{center}
             {\tenrm Fig.~\thefigure. #1}
              \end{center}}
        \fi}

\newcommand{\tcaption}[1]{
        \refstepcounter{table}
        \setbox\@tempboxa = \hbox{\tenrm Table~\thetable. #1}
        \ifdim \wd\@tempboxa > 6in
           {\begin{center}
        \parbox{6in}{\tenrm\baselineskip=12pt Table~\thetable. #1 }
            \end{center}}
        \else
             {\begin{center}
             {\tenrm Table~\thetable. #1}
              \end{center}}
        \fi}

\def\@citex[#1]#2{\if@filesw\immediate\write\@auxout
	{\string\citation{#2}}\fi
\def\@citea{}\@cite{\@for\@citeb:=#2\do
	{\@citea\def\@citea{,}\@ifundefined
	{b@\@citeb}{{\bf ?}\@warning
	{Citation `\@citeb' on page \thepage \space undefined}}
	{\csname b@\@citeb\endcsname}}}{#1}}

\newif\if@cghi
\def\cite{\@cghitrue\@ifnextchar [{\@tempswatrue
	\@citex}{\@tempswafalse\@citex[]}}
\def\citelow{\@cghifalse\@ifnextchar [{\@tempswatrue
	\@citex}{\@tempswafalse\@citex[]}}
\def\@cite#1#2{{$\null^{#1}$\if@tempswa\typeout
	{IJCGA warning: optional citation argument
	ignored: `#2'} \fi}}
\newcommand{\citeup}{\cite}

\def\fnm#1{$^{\mbox{\scriptsize #1}}$}
\def\fnt#1#2{\footnotetext{\kern-.3em
	{$^{\mbox{\sevenrm #1}}$}{#2}}}

\font\twelvebf=cmbx10 scaled\magstep 1
\font\twelverm=cmr10 scaled\magstep 1
\font\twelveit=cmti10 scaled\magstep 1
\font\elevenbfit=cmbxti10 scaled\magstephalf
\font\elevenbf=cmbx10 scaled\magstephalf
\font\elevenrm=cmr10 scaled\magstephalf
\font\elevenit=cmti10 scaled\magstephalf
\font\bfit=cmbxti10
\font\tenbf=cmbx10
\font\tenrm=cmr10
\font\tenit=cmti10
\font\ninebf=cmbx9
\font\ninerm=cmr9
\font\nineit=cmti9
\font\eightbf=cmbx8
\font\eightrm=cmr8
\font\eightit=cmti8


\centerline{\tenbf INSTANTON INTERACTIONS AND NON-PERTURBATIVE PARTICLE
PRODUCTION IN QCD \footnote{Talk presented at the Workshop `Continuous Advances
in QCD', Minneapolis, 18-20 Feb. 1994}}
\baselineskip=16pt
\vspace{0.4cm}
\centerline{\tenrm IAN BALITSKY \footnote{On leave of absence from
St.Petersburg Nuclear
Physics Institute, 188350 Gatchina, Russia}}
\baselineskip=13pt
\centerline{\tenit Physics Department, Penn State University}
\baselineskip=12pt
\centerline{\tenit State College, PA 16802}
\vspace{0.4cm}
\abstracts{I discuss the possible instanton-induced  multiparticle production
in hard processes in QCD}
\vfil
\rm\baselineskip=14pt
\section{Introduction }
\vspace*{-0.7cm}
\subsection{Instantons and  their interactions}
\vspace*{-0.35cm}
It is well known that the potential energy in gauge theories has a periodic
structure with respect to Chern-Simons number. There is an infinite set of
gauge rotated vacua and the instanton is a transition under the potential
barrier separating two adjacent vacua. As usual, quantum tunneling in a
classically forbidden zone is described by a solution of the equations of
motion in imaginary time, or, in other words, by the solution of the Euclidean
field equations. In QCD we know the explicit form of such solution - it is the
famous Belavin-Polyakov-Schwartz-Tyupkin instanton characterized by orientation
in color space $u$, position of the center $x_0$, and size $\rho$ \footnote{We
use here the notations $\sigma_\mu^{\alpha\dot\alpha} = (-i\sigma, 1)$,
$\bar\sigma_{\mu\dot\alpha\alpha} = (+i\sigma, 1)$
($\sigma$ are the standard  Pauli matrices) related to standard
t'Hooft symbols as $\sigma_{\mu}\bar{\sigma}_{\nu}={\rm g}_{\mu\nu}
+i\eta^a_{\mu\nu}\tau^a$,
$\bar{\sigma}_{\mu}\sigma_{\nu}={\rm g}_{\mu\nu}
+i\eta^a_{\mu\nu}\tau^a$. }:
\begin{equation}
A_{\mu}^I=-{i\over g} \rho_1^2{\bar{u}(\sigma_{\mu}(\bar x
-\bar{x_0})-(x-x_0)_{\mu})u\over (x-x_0)^2 ((x-x_0)^2+\rho^2)}
\end{equation}
 The action of the instanton is $S_0={8\pi^2\over g^2}$ and the exponential of
this number $e^{-S_0}$ is a semiclassical factor for the tunneling. The
propagators and determinant in the instanton background were calculated long
ago and apart from that there is not a great deal left to learn about a single
instanton - the real physics comes from the instanton interactions.

The interest to instanton interactions originates from the paper of Callan,
Dashen, and Gross \cite{cdg}, who tried to obtain confinement in a vacuum
populated by instantons. Their attempts were not sucsessful - we know pretty
well now that instantons have nothing to do with confinement - but the ideas
developed in that paper are widely used now in construction of instanton-based
models of QCD vacuum, which seem to be working reasonably well (at least if we
are asking the proper questions, see e.g. ref.\cite{ivac}). Another important
aspect of instanton interactions is that they are related to the asymptotics of
perturbative series (see ref.\cite{asym} for a review). But in recent years the
main part of instanton activity was related to another feature of instanton
interactions, namely to the fact that they determine non-perturbative particle
production.

To illustrate this property, consider the interaction of a widely separated
instanton ($I$) and antiinstanton
($\bar{I}$) that was obtained in the same paper by CDG \cite{cdg}. It is given
by the following "dipole-dipole formula":
\begin{equation}
U_{int} ~=~{32\pi^2\over g^2}
{}~\rho_1^2\rho_2^2~u_{\mu}u_{\nu}\left(4{R_{\mu} R_{\nu}\over R^6}-{1\over
R^4}\right)
\label{dipdip}
\end{equation}
where vector $u$ describes the relative $I\bar{I}$ orientation in color space.
The instanton-antiinstanton interaction is defined here simply as the
difference between the action of the  $I\bar{I}$ configuration and $2S_0$. A
very important observation is that the expression in the parentheses in the
above formula is simply the bare gluon propagator (in coordinate space). So the
picture which corresponds to Eq.(2) is the following: the instanton emits
gluons somewhere near the center $x_0^I$, after that they propagate a long
distance $R$ and finally they are absorbed by $\bar I$ near the center
$x_0^{\bar I}$, see Fig.1. 
\begin{figure}
    \begin{center}
        \begin{picture}(100,170)
        \end{picture}
    \end{center}
    \caption[xxx]{ \small
                   Instanton-antiinstanton interaction as a result of
                   exchange of gluons
                    }
\end{figure}
It is clear from Fig.1 that since the gluon lines are simply bare propagators
we can calculate the imaginary part of $e^{U_{int}}$ using the Cutkovsky rules
and thus we arrive at the conclusion that the imaginary part of $U_{int}$
determines the total cross section of instanton-induced particle production.

 Unfortunately, in QCD the integrals over instanton size are normally divergent
and therefore instanton-induced particle production was not considered to be of
interest for many years. The recent surge of interest to instanton-induced
processes was inspired by Ringwald's paper \cite{ring} where he had suggested
that in the electroweak theory instantons give us large cross sections  with
baryon number violation (BNV) at energies $\sim 10 Tev$. Although the common
belief now is that these BNV-processes probably are still too small to be
observed this has stimulated  general study of the possibility of multiparticle
production by classical field configurations in different theories. The
 present paper is devoted to particle production induced by QCD-instantons in
hard processes such as deep inelastic scattering.

\subsection{Baryon number violation due to electroweak instantons}
\vspace*{-0.35cm}

In the electoweak sector of the Standard Model there are no exact solutions of
Euclidean equations but there is an approximate solution which is called a
constrained instanton \cite{affleck}. It is a superposition of the QCD-like
gauge instanton at the distances $x-x_0\sim \rho \ll m_W^{-1}$  and of the
exponentially falling tail $\sim e^{-m_W|x-x_0|}$ at large distances $x-x_0 \gg
m_W^{-1}$  (where $m_W$ is a mass of W-boson). There is also a Higgs component
of the instanton field which leads to the additional term in the instanton
action proportional to the size of instanton. The total action of the
electroweak instanton is given by the formula
\begin{equation}
S_0~=~{8\pi^2\over g_W^2}~+~\pi^2v^2\rho^2
\end{equation}
where $v$ is the v.e.v. of Higgs field. Unlike QCD,  the integrals over
instanton size  in the electroweak theory are convergent due to the last term
in Eq.(3).
Therefore, in the electoweak theory  we must have well-defined
instanton-induced cross sections which turn out to have a fascinating property
- they lead to processes with baryon (and lepton) number violation \footnote{
Recall that in QCD, because of  the axial anomaly, tunneling under potential
barrier separating the two vacua with different Chern-Simons number changes the
axial charge by factor 1 (1 for the $I$ and -1 for the $\bar I$). The analog of
the axial charge for the electroweak instanton transition is the sum of baryon
and lepton numbers ($B+L$) so we shall see violation of baryon and lepton
number $\Delta B=\Delta L=3$ for every (electroweak) instanton transition}.
This remarkable property of the electroweak instantons was realized already in
the pioneering paper by t'Hooft \cite{pio} but in the same paper the rate of
this transition was estimated to be of the order of $e^{-4\pi/\alpha_W}\sim
10^{-170}$ (the factor in the exponent  is simply twice the gauge part!
 of the instanton action, see Eq.(
3)) and for the next 15 years the interest to this process was lost.
 Things had changed when Ringwald \cite{ring} discovered that at high energies
the BNV particle production can be enhanced by the large number of emitted W's.
His result reads:
\begin{equation}
\sigma_{BNV}~\sim ~e^{-{4\pi\over \alpha_W}(1-{9\over 8}\epsilon^{4/3})}
\label{ring}
\end{equation}
where $\epsilon=E/E_0$ and $E_0=\sqrt{6}\pi m_W/ \alpha_W \sim 17~TeV$ is the
so-called sphaleron energy. Taken literally, this result predicts the cross
section with BNV of order of 1 at energies $\sim 17Tev$! But very soon it was
realized that the above formula is valid only at small $\epsilon$. We can
parametrize the cross section with BNV in the following way:
\begin{equation}
\sigma_{BNV}~\sim ~e^{-{4\pi\over \alpha_W}F(\epsilon)}
\end{equation}
and the Ringwald's formula is the first term of the expansion of $F(\epsilon)$
in powers of $\epsilon$. Now we know a couple of further terms in this
expansion \cite{NTL,NTTL}:
\begin{equation}
F(\epsilon)=1-{9\over 8}\epsilon^{4/3}+
{9\over 16}\epsilon^2-
{3\over 32}(4-3{m_H^2\over m_W^2})\epsilon^{8/3}\ln {1\over \epsilon}+...
\end{equation}
where $m_H$ is the mass of the Higgs boson. Small $\epsilon$ correspond to a
widely separated \II\ and the expansion of $F(\epsilon)$ goes in powers of
$\epsilon^{2/3}$ since this parameter corresponds to $\rho^2/R^2$ - the usual
small parameter for a system of weakly interacting instantons. (For example,
Ringwald's formula Eq.\ref{ring} follows from the dipole-dipole approximation
Eq.(2) for the \II\ interaction). Unfortunately, we have no quantitative
description of the behavior of $F(\epsilon)$ at large $\epsilon$ since we do
not know how to describe the strong interaction of instantons at small
separations.
However, qualitative arguments based on unitarity \cite {zakh91,magg91}
demonstrate that this function is effectively bounded from below, so the
scenario of behavior of  $F(\epsilon)$ and the cross section looks like in
Fig.2.
\begin{figure}
    \begin{center}
        \begin{picture}(100,150)
        \end{picture}
    \end{center}
    \caption[xxx]{ \small
                  Scenario of behavior of the instanton-induced \crse s at
large energies. Dashed line corresponds to Ringwald's formula Eq.\ref{ring}
                    }
\end{figure}
This means that $\sigma_{BNV}$ is at best $\sim 10^{-80}$ and will never be
observed experimentally.
 After this scenario became commonly accepted the interest shifted to general
questions of multiparticle production by classical field configurations in
various models such as scalar field theory.

\subsection{QCD instantons in hard processes}
\vspace*{-0.35cm}
I believe, however, that there are well-defined and (possibly) observable
instanton-induced \crse s in QCD itself\footnote{First dicsussion of the
possibility to observe QCD-instanton was presented in ref.\cite{rys}}. Of
course, we have no such clear trigger as baryon number violation here. The
reason is that the violation of chirality can come not only from instantons but
can be due to quark vacuum condensate $\bar{\psi}\psi$ formed by large-scale
vacuum fluctuations (probably with the large-scale instantons among them). On
the other hand, the strong coupling constant is much greater than the weak one
so even if the instanton-induced \crse s are bounded by $e^{-2\pi/\alpha_s}$ as
shown in Fig.2 there is still a possibility that they can be sizeable eniugh to
observe them, especially since they are likely to produce events with a very
specific structure of the final state, and such
peculiarities may be subject to experimental search. One may wonder, however,
about the usual objection that the integrals over the instanton sizes in QCD
are divergent and therefore meaningless. Our answer is that if we consider a
$hard$ process with characteristic scale $Q^{-1}$,  the integrals over the
instanton sizes converge at the scale $\rho\sim {1\over Q\alpha_s(Q)}$ and the
instanton-induced \crse\ is well-defined.

  A common wisdom about hard processes such as deep inelastic scattering is
that due to the factorization theorems the contributions coming from small and
large distances can be separated into the coefficient functions in front of
light-cone operators and their matrix elements - parton densities. The
coefficient functions correspond to small distances and hence can be
calculated perturbatively while the  parton distributions
absorb all the information about the dynamics of large distances and
are fundamental quantities extracted from the experiment. Less widely known is
the fact that from the theoretical point
of view this picture is not complete. An indication that some
contributions may be missing comes
 from the asymptotic nature of the perturbative series.
 This series is non-Borel-summable, which means
that any attempt to attribute  a quantitative meaning to the sum of
the  series would produce an exponentially small non-perturbative
contribution to coefficient functions which comes from the
small-scale vacuum fluctuations. The explicit calculation performed
in collaboration with Dr.V.Braun (see ref.\cite{bal93a,bal93b} and Sect.4 of
present paper) confirms this:
the deep inelastic cross section indeed possesses
exponential contributions of the form
$\Phi(x)\exp[-4\pi F(x)/\alpha_s(Q^2)]$, where $\Phi(x)$, $F(x)$
are  some
functions of Bjorken $x$  \footnote{It should be emphasized that these $real$
contributions to the \crse s are not directly related to the divergence of
perturbative series for the coefficient functions which is governed by
$imaginary$ \II\ terms arising after the change of sign of coupling constant,
see ref.\cite{asym} for the discussion}. In a hard process the height of
instanton potential barrier is determined by virtuality which thus plays the
role of "sphaleron energy" $E_0$ and we shall see below that the formulas for
the QCD-instanton look similarly to the electroweak ones with the substitution
${\sqrt3\over 2}\epsilon^{2/3}\rightarrow \frac{1-x}{1+x}$. Therefore we would
like to expect the scenario of the behavior of the \crse\ as a function of
Bjorken $x$ to be as shown in Fig.3.
\begin{figure}
    \begin{center}
        \begin{picture}(100,160)
        \end{picture}
    \end{center}
    \caption[xxx]{ \small
                  Scenario of behavior of the instanton-induced \crse s in deep
inelastic scattering as a function of Bjorken $x$. Dashed line corresponds to
Eq.\ref{ring} $\sigma_I(x)=exp{-{4\pi\over \alpha_s}(1-{3\over 2}{(1-x)^2\over
(1+x)^2})}$
                    }
\end{figure}
Although the quantitative behavior of $F(x)$
at intermediate $x$ (which correspond to $\epsilon \sim 1$) is not known yet,
we may proceed as follows: start from the well-studied situation of small
$\epsilon \sim 1-x$ where not
only the instanton sizes are small, but also the $I\bar{I}$ separation
is much larger than these sizes. After that we can move
to smaller $x$ (which correspond to strongly interacting  $I$ and $\bar{I}$)
 trying to be careful to collect to semiclassical
accuracy all the dependence on $\rho^2/R^2$ in the exponent.To this
end,we shall have in mind  the valley method \cite{bal86}, in which
all the dependence on the $\bar I I$
separation is absorbed in the action $S(\xi)$ on the $\bar I I$
configuration. However, we do not take into
account corrections of order $\rho^2/R^2$ in the preexponent so our results for
\crse s are at best order-of-magnitude estimates.

We have found that instantons produce a well-defined and calculable
contribution to the cross section for deep inelastic scattering
for $x\sim 0.3-0.5$ and large $Q^2\sim 100 - 1000 GeV^2$,
which turns out, however, to be rather small ---
of order $10^{-2}-10^{-5}$ compared to the
perturbative cross section. The dominating Feynman diagrams in our calculation
correspond to
 $2\pi/\alpha(\rho_\ast)\sim 15 $ gluons and $2n_f-1 =5$ quarks
in the final state with low energy, of order $\rho_\ast^{-1}
\sim 1\,GeV$. They  are produced in a spherically symmetric way
 in the c.m. frame of the partons colliding
through the instanton and I believe that such a distribution over the momenta
in the final state which is quite different from the one arising from
perturbative analysis may be subject to experimental search.

Another interesting example of instanton-induced effects in hadron-hadron
scattering
 is a production of a single jet with large $p_{\perp}$ in the final state and
the hope is that for this
hard process the ratio signal/background for the instanton-induced \crse\
will be much better. Unfortunately, there is a theoretical price to pay for
this advantage: one should calculate the so-called hard-soft quantum
corrections - this problem was not solved during the study of baryon number
violation due to instantons. For a rough estimate we adopt a certain plausible
model for these hard-soft corrections and find that it leads to a well-defined
cointributions of the small-size instantons just as in the case of DIS.

The paper is organized as follows: in Sect.2 I explain the valley method for
the calculation of instanton-induced cross sections. In Sect.3 I will construct
the Minkowskian valley for the double functional integral describing the cross
sections. And finally, I will outline in Sect.4 the calculation of
instanton-induced contributions to the structure functions of deep inelastic
scattering and present the numerical results. Subsection 4.4 and Sect.5 are
devoted to the discussion of possible instanton-induced \crse s in other hard
processes in QCD.

\section{Valley Method}

The saddle-point gaussian approximation is the usual technique to
calculate functional integrals in weak coupling theories. The valley
method \cite{bal86} constitute a generalization to cases in which
physically relevant `approximate solutions' can be given.
The case of an instanton-antiinstanton pair at large separation is a
typical example. To illustrate the idea we consider quantum mechanics
in a double-well potential with the instanton being the simple kink
\begin{equation}
\phi_{I}~=~{1\over 2} ({\rm th}{\alpha-t \over 2}~+~1)  ~~~~~~~
\phi_{\bar I}~=~{1\over 2} ({\rm th}{t+\alpha \over 2}~+~1)
\end{equation}
describing the tunneling between the two minima. We want to calculate
the non-perturbative part of the vacuum energy
\begin{equation}
Z~=~N^{-1}~\int {\cal D}\phi ~ \exp \left\{ -{1\over g^2}~\int ~dt~
{1\over 2} [(\stackrel{.}{\phi} )^2 +\phi^2(1-\phi^2)] \right\}
\end{equation}
which at small $g^2$ is dominated by the $I\bar I$ configurations
\cite{bog}.
For infinitely large separations the $I \bar I$ configuration is just
the sum of two kinks. It obeys the classical field equations and
possesses two zero modes corresponding to the independent translations
of both instantons. The zero modes can be rediagonalized in such a way
that one of them describes the trivial translations of the complete
$I\bar I$ configuration and the other changes in the instanton
separation. For large but finite separations (as compared to the
instanton size which we chose as 1 for this toy model) the second one
becomes a quasizero mode: the action varies slowly along this direction
in functional space but grows rapidly in orthogonal directions, which
correspond to changes in the instanton profile and are associated with
normal modes. As a landscape in functional space this looks like a
steep canyon with the course of the valley corresponding to the
quasizero mode (see Fig.4).
\begin{figure}
    \begin{center}
        \begin{picture}(100,140)
        \end{picture}
    \end{center}
    \caption[xxx]{ \small
                  A picture of functional space near the valley
                  configuration                    }
\end{figure}

In order to integrate over the field configurations close to the $I
\bar I$ valley one has to perform the following steps:
(i) determine the course of the valley in the functional space,~
(ii) perform the Gaussian integrations in the directions orthogonal to
this valley, and (iii) carry out the final integration along the valley.
In step (i) one determines the valley as the trajectory in functional
space
which
minimizes the action. For any direction orthogonal to the valley the
constraint
\begin{equation}
(\phi-\phi_v, \omega(\alpha){\partial\phi_v\over \partial\alpha})=0
\end{equation}
must be fulfilled, where $(f,g)$ denotes the usual scalar product of
functions $\int dt f(t)g(t)$ and $\omega(\alpha,t)$ is a suitable
weight function. Applying the standard technique of Lagrange
multipliers this leads to the following classical constraints (the
`valley equations' \cite{bal86})
\begin{equation}
\left. {\delta S\over \delta \phi(t)}
\right|_{\phi=\phi_v}~=~\chi(\alpha)
\omega (\alpha,t) {\partial\phi_v(\alpha,t)\over \partial\alpha}
\label{val}
\end{equation}
where $\chi(\alpha)$ is the Lagrange multiplier. This equation has to
be solved with the boundary condition that for infinite $\alpha$ the valley
$\phi_v(\alpha)$ approaches the configuration of an infinitely separated
$I\bar I$ pair.
\begin{equation}
\phi_v(\alpha,t) ~\rightarrow~ \phi_I(t-\alpha)~+~ \phi_{\bar
I}(t+\alpha) ~-~1
\end{equation}
As the valley action increases monotonously with $\alpha$
\begin{equation}
{\partial S_v(\phi(\alpha))
\over  \partial \alpha} ~=~ \chi(\alpha)~ \left(
{\partial \phi_v\over \partial \alpha}, \omega
{\partial \phi_v\over \partial \alpha} \right) ~\ge~ 0
\end{equation}
we can conclude that $\alpha=0$ corresponds to the classical
perturbative vacuum and that $S$ reaches a finite value for large
$\alpha$ only if $\chi(\alpha) \rightarrow 0$, i.e. for a classical
solution (see eq. (\ref{val})).
Generally speaking the valley always interpolates between two classical
solutions, in our case the $I\bar I$ configuration and the vacuum.

The integration over the orthogonal Gaussian modes (step (ii)) is performed
using the standard Faddeev-Popov trick: we insert
\begin{equation}
1~=~ -\int d\alpha~ \delta(\phi-\phi_v,\omega \phi_v')~
1~=~ -\int d\alpha~ \delta(\phi-\phi_v,\omega \phi_v')~
\{ (\phi_v,\omega\phi_v')-
(\phi-\phi_v, (\omega \phi_v')')\}
\end{equation}
with $\phi_v'\equiv d\phi_v/d\alpha$. \footnote{Strictly speaking an additional
$\delta$-function, $\delta((\phi-\phi_v,d\phi_v/dt))$
is needed to exclude total translations from
the integral but this adds only some technical complexity without
changing the arguments \cite{bal86}}.
Then we shift $\phi$ to $\phi+\phi_v$ and expand the action in powers
of $\phi$
\begin{equation}
S(\phi+\phi_v)~=~S(\phi_v)+(\phi,L_v)+{1\over 2} (\phi,
\Box_v\phi)+O(\phi^3)
\end{equation}
where $L_v=\delta S/\delta \phi_v$ and
$\Box_v=-\partial^2+1-6\phi_v+6\phi_v^2$ is the operator of the second
derivative of the action.

Now comes the central point: the linear term $(\phi,L_v)$ in the
expansion (14) vanishes due to the factor $\delta(\phi,\phi_v')$ in
the integrand and the valley equation (\ref{val}).

Thus $\phi_v$ enters the functional integral like a classical solution
(for which $L=0$).
\begin{equation}
N^{-1}\int~d\alpha~ (\phi_v,\omega\phi_v')~e^{-{1\over g^2} S(\phi_v)}
{}~\int{\cal D}\phi~ \delta(\phi, \omega\phi_v')~e^{-{1\over
2g^2}(\phi,\Box_v \phi)}~(1+O(g^2))
\end{equation}
All effects $\sim 1/g^2$ originate from the classical action. Quantum
corrections come from the terms of order $\phi^3$ in Eq.(14) and from
the collective coordinate Jacobian (13).

Finally in step (iii) the explicite integration over the valley
parameter $\alpha$ must be performed. In this case it gives the
non-perturbative part of the vacuum energy, see ref. \cite{bal86}.

The crucial point of the whole procedure is the vanishing of the linear
term $(\phi,J)$, which will occure for any weight function
$\omega(\alpha,t)$. Hence, at first sight, any valley starting from
infinitely separated instantons and antiinstantons seems appropriate.
The fact that some choices are worse than others shows up in the size
of the quantum corrections. A `good' valley should minimize them. The
standard recipe to find such a valley is the following: start from
infinitely separated instantons and follow the direction of the
negative quasizero mode of the operator $\Box_v$ (see the discussion in
ref. \cite{valmin,NTTL}).
\begin{eqnarray}
\phi_v& ~~\stackrel{\alpha\rightarrow \infty}
{\longrightarrow}~~&
{1\over 2} {\rm th}{t+\alpha\over 2} ~-~
{1\over 2} {\rm th}{t-\alpha\over 2} \nonumber \\
\phi_v'& ~~\stackrel{\alpha\rightarrow \infty}
{\longrightarrow}~~&
\phi_- ~\sim~ {\rm ch}^{-2}{t+\alpha\over 2} ~+~
{\rm ch}^{-2}{t-\alpha\over 2}
\end{eqnarray}
All valleys satisfying Eq.(\ref{val}) are appropriate to calculate the
nonperturbative part of the vacuum energy. (Of course, the final answer
obtained after integrating over the valley parameter $\alpha$
is the same for all valleys.) The simplest choice for such a valley is
the sum of the kinks
\begin{equation}
\phi_v ~=~
{1\over 2} {\rm th}{t+\alpha\over 2} ~-~
{1\over 2} {\rm th}{t-\alpha\over 2}
\end{equation}
which trivially satisfies the condition (16) and obeys the valley
equation (\ref{val}) with the weight function
\begin{equation}
\omega(\alpha,t)~=~ {1\over 4}{ e^{\alpha} {\rm sh}
(\alpha)  \over {\rm ch}(t)
{}~{\rm ch}(\alpha) ~+1}
\end{equation}
The corresponding Lagrange multiplier is
\begin{equation}
\chi(\alpha)={12\over \xi^2} ~~~,~~~ \xi=e^{\alpha}
\end{equation}
and the valley action equals
\begin{equation}
S_v\equiv S(\phi_v)=~{6\xi^2-14\over (\xi-1/\xi)^2}~-~ {17 \over
3}~+~ \left[ {(5/\xi~-~\xi)(\xi+1/\xi)^2\over
(\xi-1/\xi)^3}~+1\right] \ln \xi
\label{act}
\end{equation}
For $\alpha=0,~\xi=1$ this expression gives zero
(perturbative vacuum) and with increasing $\alpha$ it approaches 1/3
which is twice the instanton action. Here $1/\xi$ is the
small parameter corresponding to an expansion of the $I\bar I$
interaction at large separations.
The leading terms in the expansion of the valley action are
\begin{equation}
S_v~=~ {1\over 3} ~-~ {2\over \xi} ~+~ {12\over \xi^4} \ln \xi
{}~+~ ...
\end{equation}
In the integral (15) $\alpha$ is typically of the order $\ln (-g^2)$
\footnote{The sign of $g^2$ has to be changed in order to extract the
nonperturbative part of the partition function (8), see \cite{bog}}.
such that the third term $\sim \xi^{-4}\ln \xi$ (and higher ones)
in eq.(21) mix with the quantum corrections $O(g^2)$.
If there were an
extra parameter dominating $\xi$ (e.g. the energy for the calculation
of the correlator $<\phi(E)\phi(-E)>$) the expansions in $g^2$ and
$\xi$ would be independent. (The latter is the case for high-energy
instanton-induced particle production at high energies
where $\xi \sim (E/E_0)^{2/3}$).

Starting from the double-well valley (17) it is easy to construct
a suitable valley for QCD. Due to the conformal invariance of QCD
at the tree level it is possible to construct a whole family of $I\bar
I$ configurations with finite separattion from a spherically symmetric
configurations with separation zero \cite{yung88}. It is known, that for this
spherical ansatz  (and for collinear gauge orientations) QCD is
equivalent at tree level to ordinary double-well quantum mechanics
(specified by Eq.(8)). For
\begin{equation}
A_{\mu}(x) ~=~ -{i\over g}~ (\sigma_{\mu} \bar x-x_{\mu}) ~x^{-2}~
\phi(t,\alpha)
\end{equation}
with $t=\ln x^2/\rho^2$ the QCD action coincides with the simple
quantum mechanical expression (8) up to an overall factor
$48\pi^2/g^2$. Using the quantum-mechanical valley (17)
we obtain thus the gauge field in the form
\begin{equation}
A_{\mu}(x)^v ~=~ -{i\over g}~ (\sigma_{\mu} \bar x-x_{\mu}) ~
\left( {\rho^2/\xi \over x^2+\rho^2/\xi}-
{\rho^2 \xi \over x^2+\rho^2 \xi} \right)
\end{equation}
which coincides with the sum of one instanton field in the regular
gauge with radius $\rho \sqrt{\xi}$ and one antiinstanton field in
the singular gauge with radius $\rho/\sqrt{\xi}$ up to a gauge
transformation with the matrix $\bar x/ \sqrt{x^2}$. This gauge field
obeys the valley equation
\begin{equation}
\left. {\delta S \over \delta A_{\mu}(x)}  \right|_{A=A_v}
 ~=~ \chi(\xi)
\omega(x,\xi) \xi~ {dA_v^{\nu}(x,\xi)\over
d\xi}
\label{valeq}
\end{equation}
 The weightfunction $\omega(x,\xi)$ is simply Eq.(18) taken at $t=\ln
x^2/\rho^2$.
To obtain the $I\bar I$ valley configuration for arbitrary sizes
$\rho_1,~\rho_2$ and separations $R$ one has to perform the translation
$x\rightarrow x-x_0$, the inversion $(x-a)_{\mu} \rightarrow
r^2(x-a)^{-2}(x-a)_{\mu}$
and a gauge transformation with the matrix $x(\bar x -\bar
R)R/\sqrt{x^2(x-R)^2R^2}$. After some algebra one obtains the final answer for
the gauge valley in
the form
\begin{equation}
A^v_{\mu}~=~ A^I_{\mu}~+~ A^{\bar I}_{\mu} ~+~ \tilde{A}_{\mu}
\end{equation}
where
\begin{equation}
A_{\mu}^I=-{i\over g} {\sigma_{\mu}\bar x -x_{\mu} \over x^4 \Pi_1}
\rho_1^2 ~~~,~~~
A_{\mu}^{\bar I}=-{i\over g} {R(\bar \sigma_{\mu} (x-R)
-(x-R)_{\mu})\bar R \over R^2 (x-R)^4 \Pi_2}
\rho_2^2 ~~~,~~~
\label{eval}
\end{equation}
\begin{eqnarray}
&&\tilde{A}_{\mu}= {i\over g} ~{\rho_1\rho_2\over \xi}~
\left\{  {x(\bar x-\bar R)\sigma_{\mu}
\bar x \over x^4 (x-R)^2 \Pi_1}-
{R(\bar x-\bar R)\sigma_{\mu}
\bar x (x-R)\bar R\over R^2x^2 (x-R)^4 \Pi_2} \right. \nonumber \\
 &&+{\sigma_{\mu}
\bar R\over x^2 (x-R)^2 \Pi_1\Pi_2} ~+~
\rho_1^2 \left( 1-{\rho_2\over \xi\rho_1} \right)
{\sigma_{\mu}
\bar x\over x^4 (x-R)^2 \Pi_1\Pi_2} \nonumber \\
 &+&\rho_2^2 \left( 1-{\rho_1\over \xi\rho_2} \right)
{R \bar \sigma_{\mu} (x-R)
\bar R\over R^2x^2 (x-R)^4 \Pi_1\Pi_2}
\nonumber \\
&-&
\left. {\rho_1\rho_2\over \xi}~
{R (\bar x-\bar R) \sigma_{\mu} \bar x
\over x^4 (x-R)^4 \Pi_1\Pi_2} ~-~(trace)\right\}
\label{evalad}
\end{eqnarray}
where $\Pi_1=1+\rho_1^2/x^2$ and $\Pi_2=1+\rho_2^2/(x-R)^2$.
'$O-(trace)$'
 means the traceless part of $O$. (Strictly speaking one has to
subtract ${1\over 2} {\rm Tr}O$.)
Thus the valley field is a sum of instanton and antiinstanton in a
singular gauge with relative orientation collinear with $R$ (the
maximal attractive orientation) plus a small additional field
proportional to $1/\xi$.

The valley equation for this configuration has the usual form
\begin{equation}
-{\cal D}_{\mu}G_{\mu\alpha} ~=~ 2\chi(\xi)\omega(x,\xi)\xi~
{\partial A_{\alpha}\over \partial \xi}
\label{valleyeq}
\end{equation}
where
\begin{equation}
\omega(x,\xi) ~=~ {\xi^2-1\over (x^2+\rho_1^2)^2\rho_1^{-2}
+ ((x-R)^2+\rho_2^2)^2\rho_2^{-2}} ~~~~.
\end{equation}
while $\chi(\xi)$ is given by eq.(19) and $\xi$ depends on the new variables
$\rho_1,~\rho_2$, and $R$ according to
\begin{equation}
\xi ~=~ {R^2+\rho_1^2+\rho_2^2 \over 2\rho_1\rho_2}
{}~+~\sqrt{\left( {R^2+\rho_1^2+\rho_2^2 \over 2\rho_1\rho_2}\right)^2-1}
{}~+~ \sim
{R^2+\rho_1^2+\rho_2^2 \over \rho_1\rho_2}
\end{equation}
The classical action of this configuration coincides, of course, with
Eq.(\ref{act}) up to an overall factor $48\pi^2/g^2$:
\begin{equation}
S_v~=~ {16\pi^2\over g^2}~\left( 1 ~-~ {2\over \xi} ~+~ {12\over \xi^4} \ln \xi
{}~+~ ...\right)
\label{action}
\end{equation}

 It is worth noting that the argument
of the  running coupling constant $g(\mu)$ in Eq.\ref{action} could be taken as
 $\mu=\rho_1\rho_2$
with our accuracy (since at large $R$ it should reproduce
${8\pi^2\over g^2(\rho_1)}~+~{8\pi^2\over g^2(\rho_2)}$ corresponding to
independent instantons).

\section{Valleys in Minkowski space}

In Sect.1 I explained that the imaginary part of the \II\ interaction potential
$U_{int}$ gives us the instanton-induced cross sections. However, this is true
only at relatively small energies (i.e., in the first few orders in $\epsilon$)
due to the following reasons. We use the optical theorem to relate the
 \crse\ to the imaginary part of the forward
 scattering amplitude in the instanton-antiinstanton background. It is known,
however, that in a general situation
 the physical \crse s cannot be obtained
by analytical continuation of a suitable amplitude
from the Euclidian space, which
issue is related simply to the fact that
the amplitude contains various imaginary
parts related to different physical processes. In our case,
 going over  from the  Euclidean  to the
Minkowski region and trying to find the
instanton-induced cross section as an
imaginary part of the $I\bar{I}$ amplitude,
we should pick up the contribution of the
$\langle 2|I\rN\lN\bar{I}|2\rangle$ discontinuity
and not the $\langle 2\rN\lN\II |2\rangle$ one, while
the total imaginary part obtained from Euclidean
calculations  contains both of them.
This problem does not show up at (relatively) low energies
since the `wrong' contributions only produce
quantum corrections due to the initial
and final-state interactions, but
becomes of crucial importance at high energies.

An existing technique \cite{khle91} is in principle efficient
for the evaluation of the function $F(\epsilon)$ to an arbitrary order,
but it would require complicated calculations of a variety of relevant
contributions (`quantum' corrections at the one-instanton background).
The essence of the valley approach is in a certain
resummation of relevant
terms such that the answer is given by
the classical action evaluated on the valley field trajectory, and the
`quantum' corrections only start from a
higher  order in $\epsilon^{2/3}$.
Now we will show how to generalize the valley approach in such a way
that it would become applicable to
evaluation of effects induced by classical field
configurations in particular
discontinuities of the amplitude.

A suitable  operator formalism has been
proposed by V.Braun and myself in connection with
calculations within the perturbation theory
of higher-twist effects in inclusive
particle production in \eea\  \cite{balit91}. Later we generalized this
approach for the case of diagrams in external fields such as the \II\ pair.
It turns to be convenient to encorporate ideas from
 nonequilibrium statistical physics and the
Keldysh diagram technique \cite{keld64,chou85}
in particular.  Within
this approach it is possible to trace which fields stand in
the amplitude to the right of the cut
(they are labeled as $(+)$ fields) and which ones appear to be
 to the left (so-called $(-)$
fields). This doubling of species of fields
in the functional integral allows one to
calculate a particular discontinuity
in the  amplitude  in the operator
language.  The \mael s of field operators are
given by (perturbative) diagrams
with internal lines of different type:
the $(++)$  Green functions are the ordinary Feynman
propagators with singularities of the type $-p^2-\ie$,
the $(--)$ propagators possess  complex conjugated
singularities $-p^2+\ie$,
 while the $(-+)$ propagators equal $2\pi\delta(p^{2})\theta(p_{0})$.
(The corresponding singularities in the
coordinate space are $-x^{2}+\ie,-x^{2}
-\ie$,  and $-x^{2}+\ie x_{0}$, respectively).
We argue  that classical  fields should have the same structure
of singularities in Minkowski space
as the quantum ones : the singularities of the instanton
field to the right of the cut have  the form $ \rho^{2}-x^{2}+\ie$,
the antiinstanton to the left of the cut yields   $  \rho^{2}-x^{2}-\ie$.
 The summation over all intermediate states
in the calculation of the \crse can be done implicitly
and actually induces a common boundary condition
in both the $(+)$ and $(-)$ functional integrals
(standing for the direct and the final
amplitudes, respectively). Owing to
this bounary condition the antiinstanton
field of the type $\rho^{2}-x^{2}-\ie 0$  participates in the `mirror'
right-hand
side functional integral over the $(+)$ fields. The counterpart
`mirror' instanton configuration in the $(-)$
functional integral
has the conjugate singularities  of type
$\rho^{2}-x^{2}+\ieo$ . The structure of
singularities is a direct consequence of the standard vacuum boundary
 conditions (no ingoing waves) at time $t\rightarrow - \infty$ in both
the $(+)$ and $(-)$ functional
integrals, while at $t\rightarrow +\infty$
the outgoing waves exist and
correspond to real particles produced in
the intermediate state (at the cut).
The summation over intermediate states implies that
 the boundary conditions for the $(+)$ and $(-)$ functional
 integrals at $t\rightarrow +\infty$ should coincide.
Thus, we arrive to a pair of the `original'
and the 'mirror' instantons (the \II\ pair,
to be precise) for which we take
 the valley configuration with the measure
 obtained by analytical continuation
from Euclidian space. We expand around
 the two-component valley
 field in the double ($(+)$ and $(-)$) functional integral
with the following analytical structure:
\begin{eqnarray}
         A_{+}^v&=&(\rho_{2}^{2}-x^{2}+\ie)^{-1}+
                       (\rho_{1}^{2}-(x-R)^{2}-\ie (x-R)_{0})^{-1},
             \nonumber\\
         A_{-}^v&=&(\rho_{2}^{2}-x^{2}+\ieo )^{-1}+
                      (\rho_{1}^{2}-(x-R)^{2}-\ie)^{-1}.
\label{1a}
\end{eqnarray}
Here R is the \II\ separation. Of course, eq.(\ref{1a})
is a  schematical one  -- we have displayed the structure of
singularities only and omitted all the color and spin factors.
It is easy to see that
 both the fields have indeed no ingoing
 waves at time going to minus infinity,
and coincide at large positive times.
   Thus, we  calculate the  forward
scattering amplitude at the background of the conformal
valley configuration of the above type.
So far as  we do not take into account  the
 initial-state
interactions, there is only one discontinuity
(one real gluon cannot decay into several ones),
and we reproduce by this way the result of the usual `Euclidean'
calculation \cite{khoze} at  the background of the valley field
\begin{equation}
  A_{{\rm Eucl}}^v=(\rho_{2}^{2}+x^{2})^{-1}+
  (\rho_{1}^{2}+(x-R)^{2})^{-1}.
\label{1b}
\end{equation}
(Again, only the structure of poles is shown).

The formal derivation proceed as follows. Let us consider the instanton-induced
\crse\ of the gluon-gluon scattering shown in Fig.5.
\begin{figure}
    \begin{center}
        \begin{picture}(100,140)
        \end{picture}
    \end{center}
    \caption[xxx]{ \small
                   The instanton-induced contribution to the
                    gluon-gluon scattering. Wavy lines are (nonperturbative)
                    gluons and solid lines are
                   quark zero modes (only one quark flavor is depicted).
            }
\end{figure}
This cross section can be represented as follows:
\begin{equation}
  \sum_{N} \lvac\tilde{T}\{A_{\mu}(q)A_{\nu}(p)e^{iH_{I}T}\}\rN
            \lN T\{A_{\alpha}(-q)A_{\beta}(-p)e^{-iH_{I}T}\}\rvac
\label{2d}
\end{equation}
where  $\Sigma \rN\lN$ means the summation over all
intermediate states (the partial \crse s $2\rightarrow N$).
Here $\tilde{T}$ denotes the T-product with inverse time ordering in the
complex conjugated amplitude. We are going to to rewrite this \crse\ in a form
of the
functional integral with the doubling
of species of the fields instead of
the usual treatment as the product
of two functional integrals \cite{khle91}.
The effect of this procedure will be in that the summation over all
intermediate states  is made implicit
and is replaced by imposing specific
boundary conditions on the fields.
\begin{equation}
\int\!\! DA^{-}DA^{+}D\Psi^{-}D{\Psi}^{+}\,
e^{-iS^{-}+iS^{+}} A^-_{\mu}(q)A^-_{\nu}(p)A_{\alpha}^+(-q)A_{\beta}^+(-p),
\label{dfin}
\end{equation}
where we have introduced the labels $(-)$ and $(+)$ to
distinguish the fields
 from direct and final functional integrals, respectively. (The color indices
are not shown for brevity).
We are going to treat the expression in (\ref{2d}) as {\em one} functional
integral but with a doubled number of fields and must specify to this end
the corresponding boundary conditions.
 As usually, the vacuum initial state implies
that  no ingoing
particle waves exist at $t_{i}\rightarrow -\infty$.
Hence the $(+)$ fields should have only negative
frequencies at  time going to minus infinity,
while the $(-)$ fields contain in this limit only fields with
positive frequencies (the difference being due to
an  opposite sign in front of the $(-)$ action in Eq.(\ref{dfin})).
The only nontrivial  point are the
 boundary conditions at large positive times.
The summation over all intermediate states implies
 that that the $(+)$ and $(-)$ fields
should coincide at $t=t_{f}\rightarrow\infty$.
More accurately, the unity operator
which we have written down as the sum
 over all the Fock states $\rN\lN$ can
 equivalently be rewritten as the sum over
all intermediate states in the so-called `coordinate representation'
$|A(x)\rangle\langle A(x)|$ where $|A(x)\rangle$ are the eigenvectors
of the field operator
$ A(x)$ with the eigenvalues equal to the classical field $A(x)$
 in close analogy with ordinary quantum mechanics.
It is obvious that in the integrand of (\ref{dfin}) one should take
$A^{-}(\vec{x},t_{f})=A^{+}(\vec{x},t_{f})=A(\vec{x})$.
In other words, the boundary conditions at time equal to plus infinity
 are such that the $(+)$ and $(-)$ fields should coincide.

 We are going to describe a procedure of
 calculation of the instanton-induced \crse\ by some quasiclassical evaluation
of the functional integral
in (\ref{dfin}). As usual, saying `instanton induced', we have in mind that
the shift of integration variables is made
\begin{eqnarray}
        A_{\mu}^- &\rightarrow & A^{\bar{I}-}_{\mu}+B_{\mu}^-,
              \nonumber\\
        A_{\mu}^+ &\rightarrow & A^{I+}_{\mu}+B_{\mu}^+,
\label{2l}
\end{eqnarray}
where $A^{\bar{I}-}_{\mu}$ and $A^{I+}_{\mu}$
are the instanton field in direct,
and the antiinstanton field in the final amplitude,
 respectively, rotated to
the Minkowsky space\footnote{
The notations in Minkowski space are $\sigma^{\mu}=(1,\vec{\sigma})$,
$\bar{\sigma}^{\mu}=(1,-\vec{\sigma})$ (cf. Eq.1) and the relation to t'Hooft
symbols is $\sigma_{\mu}\bar{\sigma}_{\nu}={\rm g}_{\mu\nu}
-i\eta^a_{\mu\nu}\tau^a$,
$\bar{\sigma}_{\mu}\sigma_{\nu}={\rm g}_{\mu\nu}
-i\eta^a_{\mu\nu}\tau^a$. }
\begin{eqnarray}
   A_{\mu}^{I+}&=&\frac{i\rho_2^2}{g}
              \frac{\sigma_{\mu}\bar{x}-x_{\mu}}
               {(-x^2+\ie)(\rho_2^2-x^2+\ie)},
                                                          \label{2m}\\
   A_{\mu}^{\bar{I}-}&=&\frac{i\rho_1^2}{g} u
              \frac{\bar{\sigma}_{\mu}(x-R)-(x-R)_{\mu}}
               {(-(x-R)^2-\ie)(\rho_1^2-(x-R)^2)-\ie)} \bar{u},
   \label{2n}
\end{eqnarray}
where $u$ is the unitary matrix of the \II\ relative orientation.
The remaining `quantum' fields $B_{\mu}$ describe creation of particles
in presence of classical instanton fields $B_{\mu}$.
To have a quasiclassical expansion we would like to deal with quantum
fields $B$ of the order of unity at the background of large classical
fields $\sim 1/{\rm g}$. However, as discussed
in detail in ref.\cite{khle91},
the situation is not that simple. In our language the problem is to satisfy
the boundary condition
 $A^{\bar{I}-}_{\mu}+B_{\mu}^-= A^{I+}_{\mu}+B_{\mu}^+$ at $t=t_f$.
We easily see that the quantum fields $B_{\pm}$ {\rm defined} in
eq.(\ref{2l}) appear to be of the order of $1/{\rm g}$ at least at
 $t\rightarrow t_f$. In the technique of ref.\cite{khle91} the
constraint $A^-(\vec{x},t=t_f)=A^+(\vec{x},t=t_f)$ is removed at
a price of an additional integration over
creation and annihilation operators of the
coherent states, see below. Thus the so-called `$\Re$-term'
comes into the game which is given by
the product of the classical and quantum fields at $t=t_f$,
times $1/{\rm g}$.

Another possibility is to require
that the boundary conditions at $t=t_f$ are satisfied for both the
classical and the quantum fields separately \cite{valmin,arnmat}. To assure
this property,
 we make an additional shift of integration variables, and extract
`classical' pieces from the quantum fields $B_{\mu}^-$ and $B_{\mu}^+$.
Thus, instead of Eq.\ref{2l} we write down
\begin{eqnarray}
 A_{\mu}^- &\rightarrow &
A^{\bar{I}-}_{\mu}+\tilde{A}_{\mu}^-+B_{\mu}^-,
              \nonumber\\
        A_{\mu}^+ &\rightarrow &
        A^{I+}_{\mu}+\tilde{A}_{\mu}^++B_{\mu}^+,
\label{2o}
\end{eqnarray}
and require that
$A^{\bar{I}-}(\vec{x},t=t_f)+\tilde{A}^-(\vec{x},t=t_f)
=A^{I+}(\vec{x},t=t_f)+\tilde{A}^+(\vec{x},t=t_f)$, and
$B_{\mu}^-(\vec{x},t=t_f)=B_{\mu}^+(\vec{x},t=t_f)$.
{}From the above discussion of the bare propagators it is clear that
the simplest choice of `mirror' fields $\tilde{A}^{-}$ and $\tilde{A}^{+}$
in order to satisfy
these constraints is to take them in form
of the antiinstanton and  the instanton
field configurations in Eq.(\ref{2m}),(\ref{2n}),
 respectively, but change the
prescription to go around the singularities to
\begin{eqnarray}
\tilde{A}_{\mu}^+&=&\tilde{A}_{\mu}^{\bar{I}+}=
A^{\bar{I}-}_{\mu}(\ie\rightarrow \ie (x_0-R_0)),
       \nonumber\\
\tilde{A}_{\mu}^-&=&\tilde{A}_{\mu}^{I-}=
A^{I+}_{\mu}(\ie\rightarrow \ie x_0)).
\label{2p}
\end{eqnarray}
However, similar as  the word `quantum'
was not accurate  in application
to the field $B_{\mu}^{\pm}$ in Eq.(\ref{2l})
owing to large contributions at
$t=t_f$, now the word `classical'   becomes ambiguous in
application to the field $A^I+\tilde{A}^{\bar{I}}$,
 since the latter no longer satisfies the equations of motion.
We have  eliminated the $\Re$-term at the boundary $t=t_f$, but
only at a price of having instead linear terms in the action
inside both the $(+)$ and $(-)$ sectors.
Fortunately, this second type of linear terms is known for a long while in
connection with the problem of \II\ interaction, and can be treated by
the valley method of ref.\cite{bal86}. So, the idea is that choosing the
mirror field configuration in form of a slightly modified instanton field
we can minimize this linear term in some sence.

Now it is easy to construct the valley configuration for the
double functional integral Eq.(\ref{dfin})
corresponding to the Euclidean valley
Eq.(\ref{eval}) and satisfying the boundary conditions specified above:
no ingoing
waves at $t=-\infty$ in both the $(+)$ and $(-)$ integrals,
 and coinciding outgoing waves at
$t\rightarrow +\infty$.
As discussed above, the relevant configuration  consists of
pairs of `original' and `mirror' instantons in each sector.
Since the Euclidean valley
Eq.(\ref{eval}) is given by the sum of  $I$ and $\bar{I}$ with the
small addition (\ref{evalad}) needed to satisfy the valley
equation, we apply  the rule Eq.(\ref{2p}) to obtain the Keldysh-type
valley solution in form \footnote{The quark components of the valley can be
obtained in a similar way but since quark exchanges affect only the
preexponential behavior and we agreed to take it into account only in the
leading order in $\rho^2/R^2$ the quark fields can be taken simply as zero
modes for $I$ and $\bar{I}$.}:
\begin{eqnarray}
  A_{\mu}^{v-}&=&\frac{i\rho_1^2}{g}
              \frac{R(\bar{\sigma}_{\mu}(x-R)-(x-R)_{\mu})\bar{R}}
               {R^2(-(x-R)^2-\ie)(\rho_1^2-(x-R)^2)-\ie)} +\nonumber\\
           &+&\frac{i\rho_2^2}{g}
              \frac{\sigma_{\mu}\bar{x}-x_{\mu}}
            {(-x^2+\ie x_0)(\rho_2^2-x^2+\ie x_0)} +\tilde{A}_{\mu}^-
\nonumber\\
  A_{\mu}^{v+}&=&\frac{i\rho_1^2}{g}
              \frac{R(\bar{\sigma}_{\mu}(x-R)-(x-R)_{\mu}\bar{R}}
     {R^2(-(x-R)^2-\ie(x-R)_0) (\rho_1^2-(x-R)^2-\ie(x-R)_0)}\nonumber\\
            &+&\frac{i\rho_2^2}{g}
              \frac{\sigma_{\mu}\bar{x}-x_{\mu}}
               {(-x^2+\ie)(\rho_2^2-x^2+\ie)} +\tilde{A}_{\mu}^+
\label{mval}
\end{eqnarray}

where the small fields $\tilde{A}_{\mu}^-$ and $\tilde{A}_{\mu}^+$ are obtained
from Eq.(\ref{eval}) in a similar way.
It is easy to see that the valley solutions in Eq.(\ref{mval}) indeed satisfy
the proper  boundary conditions.

Now I'll demonstrate that the double functional integral for the Minkowskian
valley Eq.(\ref{mval}) corresponds in the leading semiclassical approximation
to the imaginary part of analytically continued Euclidean answer for valley
Eq.(\ref{eval}).
Similar as in case of the Euclidian valley in Eq.(\ref{valleyeq})
 the linear term
$D_{\mu}^-G_{\mu\nu}^-$ vanishes thanks to the $\delta(
A^--A^-_{\rm cl},D_{\mu}^-G_{\mu\nu}^-)$ factor inserted as a
Faddeev--Popov constraint. (Of course, the linear term built of
$(+)$ fields also disappears). Hence, to evaluate
the functional integral in (\ref{dfin})  to the semiclassical accuracy
we should insert  the classical valley fields
$  A_{\mu}^{v-}$ and $ A_{\mu}^{v+}$ (and take into account the
relevant determinants at the valley field background).
The action at the Keldysh-type valley configuration in
Eq.(\ref{mval}) can be evaluated by a straightforward calculation.
After some algebra we get
\begin{eqnarray}
\lefteqn{
     \frac{i}{2}\int dx\, G_{\mu\nu}^{v-}G_{\mu\nu}^{v-}
  - \frac{i}{2} \int dx\, G_{\mu\nu}^{v+}G_{\mu\nu}^{v+}=}
                                      \nonumber\\
&&\mbox{}=     \frac{i}{2}\int dx
\left\{
\frac{\rho_1^2}{(\rho_1^2-(x-R)^2)-\ie)^2}
\left(
\frac{\rho_1}{\rho_1^2-(x-R)^2)-\ie}-
\frac{\rho_2/\xi}{\rho_2^2-x^2+\ie x_0}
          \right)^2
           \right.
                                      \nonumber\\
&&\mbox{}\hspace*{1cm}+
\left.
\frac{\rho_2^2}{(\rho_2^2-x^2+\ie x_0)^2}
\left(
\frac{\rho_2}{\rho_2^2-x^2+\ie x_0}-
\frac{\rho_1/\xi}{\rho_1^2-(x-R)^2)-\ie}
          \right)^2
           \right\}
                                      \nonumber\\
&&\mbox{}-
 \frac{i}{2}\int dx
\left\{
\frac{\rho_1^2}{(\rho_1^2-(x-R)^2)-\ie (x-R)_0)^2}
           \right.
                                      \nonumber\\
&&\mbox{}\hspace*{3cm}\times
\left(
\frac{\rho_1}{\rho_1^2-(x-R)^2)-\ie (x-R)_0}-
\frac{\rho_2/\xi}{\rho_2^2-x^2+\ie }
          \right)^2
                                      \nonumber\\
&&\mbox{}\hspace*{1cm}+
\left.
\frac{\rho_2^2}{(\rho_2^2-x^2+\ie )^2}
\left(
\frac{\rho_2}{\rho_2^2-x^2+\ie }-
\frac{\rho_1/\xi}{\rho_1^2-(x-R)^2)-\ie (x-R)_0}
          \right)^2
           \right\}
                                      \nonumber\\
&&\mbox{}=-S^v_{\rm Eucl}
\left(R^2\rightarrow -R^2+\ie R_0\right).
\label{4g}
\end{eqnarray}
Thus we reobtain the answer (\ref{action}) for the Euclidian valley with the
substitution $R^2_{\rm Eucl}\rightarrow -R^2_{\rm Mink}+\ie R_0$
which corresponds to taking the imaginary part with respect to $(p+k)^2$
after the Fourier transformation.

\section{Instanton-induced contributions to structure functions of deep
inelastic scattering}
\vspace*{-0.7cm}
\subsection{Toy example: gluon-gluon DIS}
\vspace*{-0.35cm}
At first we consider the instanton-induced contribution to the deep inelastic
scattering of a virtual gluon from the real one. This process is not directly
physically relevant but it serves as a good toy model. It is known that for the
high-energy scattering the relevant gauge-invariant operator describing the
hard gluon is \cite{boss}
\begin{equation}
G_{\mu}(x)~=~e_{\alpha}  [x+\infty e,x]G_{\mu\alpha}(x) [x,x+\infty e]
\label{bos}
\end{equation}
where
\begin{equation}
 [x,x+\infty e]  =P\!\exp\left\{ i\int_0^{\infty}
                 \!d\lambda\, e_{\mu}A_{\mu} (x+\lambda e)\right\}
\end{equation}
is a gauge factor ordered along the lightlike line in the direction of
$e={q\over 2pq}-{q^2\over (2pq)^2}p$. Thus, one needs
to evaluate the contribution to the functional integral (\ref{dfin}) (with the
substitution $A_{\mu}(q)\rightarrow G_{\mu}(q), A_{\nu}(-q)\rightarrow
G_{\nu}(-q)$)
coming from the vicinity of the instanton-antiinstanton valley
configuration (and amputate external gluon legs afterwards).
Each hard gluon is substituted by the Fourier
transform of the operator (\ref{bos})
at large momentum, and brings in
the factor
\begin{equation}
   G_{\mu}^I(q)  \simeq
    \frac{1}{\rm g}\bar {q} (\sigma_{\mu}\bar{e} - e_{\mu})q
    (2\pi)^{5/2}
\frac{\rho^2}{8Q^2}(\rho  Q)^{1/2} e^{-\rho Q}
\label{1i}
\end{equation}
The cross section is given by the following
integral over the sizes of the instanton and antiinstanton
$\rho_1$ and $\rho_2$ and over their separation $R$ in the c.m.
frame \cite{khoze,bal93a,bal93b}:
\begin{equation}
\int d\rho_1\,d\rho_2\int dR_0\,\exp\left\{-Q(\rho_1+\rho_2)+iER_0-
\frac{4\pi}{\alpha_s(\rho)}S(\xi)\right\}.
\label{1q}
\end{equation}
Three important ingredients in this expression are: the factors
$e^{-Q\rho_1}$ and $e^{-Q\rho_2}$ , which come from the two hard gluon fields,
the factor $e^{iER_0}$, which is obtained from the standard
exponential factor $e^(p+q)R$ in the CM frame ($E=p_0+q_0$), and the action
 $S(\xi)$  evaluated on the instanton-antiinstanton
configuration (\ref{mval}) (recall that due to the Eq.(\ref{4g}) this action is
given by Euclidean formula (\ref{action}) with the substitution
 $\xi\rightarrow \frac{\rho_1^2+\rho_2^2 -R^2+\ie R_0}{\rho_1\rho_2}$).
  The contour of integration over $R_0$ in eq.(\ref{1q}) is going
`in the Minkowski region' along the real axis as shown in Fig.6.
\begin{figure}
    \begin{center}
        \begin{picture}(100,130)
        \end{picture}
    \end{center}
    \caption[xxx]{ \small
                  Contour of integration in the $R_0$ plane in Eq.(\ref{1q})
  }
\end{figure}
However, the integral
can be calculated by the steepest descent method
with the saddle point lying
`in the Euclidean region' on the imaginary axis \footnote{This corresponds to
 the 'Euclidean' calculation when we consider the gluon-gluon scattering in the
 \II\ background in the Euclidean region, then make the analytical continuation
 to Minkowski space by changing $E\rightarrow iE$, and take the imaginary part
(see the discussion in previous Section).}.
At small energies this saddle point is fixed by the first
dipole-dipole interaction term in $S(\xi)$.
 The saddle-point equations take the form \cite{bal93a}
\begin{eqnarray}
 Q\rho_\ast &=&
\frac{4\pi}{\alpha_s(\rho_\ast)}(\xi_\ast-2)S'(\xi_\ast)
+bS(\xi_\ast)\,,
        \nonumber\\
E \rho_\ast=  &=&\frac{8\pi}{\alpha_s( \rho_\ast)}
\sqrt{\xi_\ast-2}\,S'(\xi_\ast)\,,
\label{saddleeq}
\end{eqnarray}
where $S'(\xi)$ is the derivative of $S(\xi)$ over $\xi$, and $\rho_{\ast
1}=\rho_{\ast 2}=\rho_\ast$ and $\xi_\ast$ are the saddle-point values for the
instanton sizes and the conformal parameter, respectively.

Neglecting in (\ref{saddleeq}) the terms proportional to
$b=(11/3) N_c - (2/3)n_f$, which come from the differentiation
of the running coupling and only produce a small correction,
one finds
\begin{eqnarray}
      \xi_\ast & = & 2 - \frac{R_\ast^2}{\rho^2_\ast}
        = 2 \frac {1+x}{1-x} ,
\nonumber\\
     Q\rho_\ast &=& \frac{4\pi}{\alpha_s(\rho_\ast)}
     \frac{12}{\xi_\ast^2}
\label{saddle}
\end{eqnarray}
A numerical solution of the saddle-point equations in (\ref{saddleeq})
for the particular expression of the action $S(\xi)$ corresponding
to the conformal instanton-antiinstanton
valley is shown in Fig.7.

\begin{figure}
    \begin{center}
          \begin{picture}(100,240)
          \end{picture}
    \end{center}
    \caption[xxx]{ \small
                  The non-perturbative scale in deep inelastic
                   scattering  (instanton size $\rho_\ast^{-1}$),
                  corresponding to the solution of
                  equation Eq.\ref{saddleeq} as a
                  function of $Q$ and for $S(\xi_\ast)\sim 0.5-0.6$
                  ($\xi_\ast\sim 3-4$).
 }
\end{figure}

Note that the difference between the hard scale $Q^2$ and the
effective scale for nonperturbative effects $\rho^{-2}_\ast$ is
numerically very large.  This is a new situation compared to
 calculations of instanton-induced
 contributions to two-point correlation functions, see
e.g. refs \cite{andrei,NSVZ80,DS80}, where the
size of the instanton is of order of the large virtuality, and
indicates that the instanton-induced contributions to deep
inelastic scattering may turn out to be nonnegligible even at
values
$Q^2 \sim 1000 GeV^2$, which are conventionally
considered as a safe domain for perturbative QCD.

In the case of hard gluon-gluon scattering it is easy to collect
all the preexponential factors (to the semiclassical accuracy).
The result for the scattering of a transversely polarized hard
gluon from a soft gluon reads (cf. ref\cite{bal93a}):
\begin{eqnarray}
2E^2\sigma_{\perp} & =&
\frac{1}{6} d^2 \frac{(1-x)^2+x^2}{(1+x)^2}
\pi^{13/2} \left(\frac{2\pi}{\alpha(\rho_\ast)}\right)^{23/2}
\exp \left[ -
\left(\frac{4\pi}{\alpha_s(\rho_\ast)}+2b\right)
S(\xi_\ast )\right]\,.
\label{gluon}
\end{eqnarray}
 It is expressed in terms of the saddle-point values of
  $\rho$ and $\xi$.
Here $d\simeq 0.00363$ (for $n_f=3$) is a constant which enters
the expression for the instanton density
\begin{equation}
 d  =  \frac{1}{2}C_1 \exp[n_f C_3-N_c C_2)],
\label{d}
\end{equation}
$C_1 = 0.466, C_2 = 1.54, C_3 = 0.153$ in the $\overline{MS}$
scheme.

Note that the preexponential factor in (\ref{gluon}) is
calculated to the leading semiclassical accuracy, i.e. in the
limit $ x\rightarrow 1$, while the exponential factor
$\exp\{-(4\pi/\alpha_s +2b)S(\xi_\ast)\}$ includes all so-called
"soft-soft" corrections \cite{matt92}, arising from the
particle interaction in the final state. Thus, to the
exponential accuracy, the result given in (\ref{gluon})
is correct in the wide region of $x$, in which
 the $ \bar I I $
action differs by a finite amount from the value $S(\xi
\rightarrow\infty) = 1$  for infinitely separated instanton and
antiinstanton.
As we have discussed, the plausible scenario  is that the function $S(\xi)$
 decreases up
to values of order 1/2 as shown in Fig.3, after
which the further decrease and the corresponding rise of the
cross section is stopped by multiinstanton contributions,
see the reviews \cite{matt92}. It is reasonable to
take $S(\xi_\ast) =1/2$ as the lower bound for the
applicability of (\ref{gluon}), which translates to the
condition that the value of Bjorken $x$ should be not too
small, $x>0.25-0.3$.
 With this restriction,
and at $\alpha_s(\rho_\ast) \simeq 0.3-0.4$,
the expression on the r.h.s. of (\ref{gluon}) is of order
$10^{-2}-10^{0}$,
which means that at $Q^2 \sim 100-1000 GeV^2$ and $x< 0.25-0.40$
the nonperturbative contribution is significant.

\subsection{Instanton contributions to the structure function of a gluon}
\vspace*{-0.35cm}
For simplicity , in this Section we will obtain the instanton-induced
contribution to the cross section of DIS
from a real gluon using the optical theorem (as discussed above, it is
legitimate up to initial-state corrections). We will calculate the
 amplitude of gluon-photon forward scattering in Euclidean region,
rotate to Minkowski energies and take the imaginary part.
 As I mentioned above, we are trying to be accurate to collect all the
dependence on $\rho^2/R^2$ in
the exponent (by valley method) but we do not take into
account corrections of order $\rho^2/R^2$ in the preexponent. To this
 accuracy we need only the first nontrivial term in the
cluster expansion of the quark propagator at the $\bar I I$
background \cite{andrei}:
 $
  \langle x|\nabla_{I\bar I}^{-2}\bar \nabla_{I\bar I}|0\rangle
 =
 \int dz\, \langle x|\nabla_1^{-2} \bar\nabla_1|z\rangle
 \sigma_\xi \frac{\partial}{\partial z_\xi}\langle z|
 \bar\nabla_2 \nabla_2^{-2} |0\rangle
 $.

 The leading contribution to the gluon matrix
element of the T-product of the electromagnetic currents  is given
 by the following expression
\begin{eqnarray}
\lefteqn{
\int dz e^{iqz}
\langle A^a(p),\lambda |T\{ j_\mu(z) j_\nu (0)\}|A^a(p),\lambda\rangle
^{I\bar{I}}         =}
\nonumber\\
 &&\mbox{} = \sum_q e^2_q
 \int dU
\int \frac{d\rho_1}{\rho_1^5} d(\rho_1)
\int \frac{d\rho_2}{\rho_2^5} d(\rho_2)
\int dR \int dT \int dz e^{iqz}
\nonumber\\
&&\mbox{}\times
\frac{1}{8}
\,\,\mbox{lim}_{p^2\rightarrow 0}\, p^4
\epsilon^\lambda_\alpha \epsilon^\lambda_\beta
Tr\left\{A_{\alpha}^{\bar{I}}(p) A_{\beta}^I(-p)\right\}
e^{-\frac{4\pi}{\alpha_s} S_{\bar I I}}
\nonumber\\
 &&\mbox{}\times
 (a^\dagger a)^{n_f-1}
\left\{ a  \bar \phi_0(0)\bar\sigma_\nu
\langle 0|\nabla_2^{-2}\nabla_2\bar\partial \nabla_1 \nabla_1^{-2}
|z\rangle \bar\sigma_\mu \phi_0(z)
\right.
 \nonumber\\
 &&\mbox{} \left.
 +a^\dagger
\bar \kappa_0(z) \sigma_\mu
 \langle z|\nabla_1^{-2} \bar
 \nabla_1\partial \bar\nabla_2 \nabla_2^{-2}
|0\rangle \sigma_\nu \kappa_0(0)
+(\mu \leftrightarrow \nu, z \leftrightarrow 0)  +\ldots
\right\}
\label{formula1}
\end{eqnarray}
which corresponds to the diagram shown in Fig.8a.
\begin{figure}
    \begin{center}
        \begin{picture}(100,160)
        \end{picture}
    \end{center}
    \caption[xxx]{ \small
                   The instanton-induced contribution to the
                    structure function of a gluon (a) and of a quark
                   (b,c). Wavy lines are (nonperturbative) gluons.
                    Solid lines are
                   quark zero modes in the case that they are
                    ending at the instanton (antiinstanton),
                    and quark propagators at the $\bar I I$ background
                    otherwise.
                    }
\end{figure}
The full expression contains many more terms \cite{bal93b}
which are not shown because we have found that all of them are
of order $O(\alpha_s(Q^2))$ compared to the expression in
Eq.(\ref{formula1}).
The subscript '1' refers to the antiinstanton
with the size $\rho_1$ and the position of the center
 $x_{\bar I} = R+T$,
and the subscript '2'  refers to the instanton with the size
$\rho_2$ and the center at $x_I = T$.
 We use   conventional notations $\nabla = \nabla_\mu\sigma_\mu$ and
$\bar\nabla = \nabla_\mu\bar\sigma_\mu$, etc,  where
$\sigma_\mu^{\alpha\dot\alpha}$ and
$\bar\sigma_{\mu\dot\alpha\alpha}$ are defined in Sect.1.1. Also, we write the
quark zero modes in terms of the two-component Weil spinors
 $\psi_0 = \left(
 \begin{array}{c} \kappa_0\\ \phi_0 \end{array}\right)$,
$\psi^\dagger_0 =
\left( \bar \phi_0\,\, \bar\kappa_0 \right)$,
and $a$ and $a^\dagger$ denote the overlap integrals
 $  a = - \int~dx~ (\bar\kappa \partial \kappa),
 a^\dagger  = - \int~dx~(\bar\phi\bar\partial\phi)$.
We shall see below that
 the leading contribution
in the strong coupling comes from the following regions of integration:
\begin{eqnarray}
      z^2 &\sim & 1/(Q^2\alpha_s)
 \nonumber\\
     (z-R-T)^2+\rho_1^2 &\sim &  T^2 +\rho_2^2\sim z^2
\nonumber\\
    (z-R-T)^2 &\sim & T^2 \sim  R^2 \sim \rho_1^2\sim
\rho_2^2 \sim  z^2/\alpha_s
\label{xap}
\end{eqnarray}
and additionally
$\rho^2/R^2 \sim 1-x$
when $x$ is close to $1$.
Note that these regions of integration correspond to imaginary part of
the  $I\bar{I}$ contribution so effectively the $z_i^2$ are negative.

Since $z^2$ is small we can use the lightcone expansion (see e.g.
\cite{lk}) for the quark
propagator in the $I\bar I$ background.Using the explicit expressions
for the propagators from \cite{brown} we find:
\begin{eqnarray}
&\bar \kappa_0(x) \sigma_\mu
 \langle x|\nabla_1^{-2} \bar
 \nabla_1\partial \bar\nabla_2 \nabla_2^{-2}
|0\rangle \sigma_\nu \kappa_0(0) =\nonumber\\
& =
-\frac{1}{2\pi^4} \int_0^1 d\gamma \frac{(\rho_1\rho_2)^{3/2}}
{[(z-R-T)^2+\rho_1^2]^2[T^2+\rho_2^2]^2}
\frac{1}{\sqrt{R^2}} \mbox{Tr}
\left\{
\frac{\bar\sigma_\nu z \bar\sigma_\mu}
     {z^4}\left[(z-R-T)
\right.\right.
\\
 & +
\left.\left.
\rho_1^2\frac{(\gamma z-R-T)}{(\gamma z-R-T)^2}\right]
\bar R \frac{1}{\sqrt{1+\rho^2_1/(\gamma z-R-T)^2}}
 \frac{\partial}{\partial \gamma}
 \frac{1}{\sqrt{1+\rho^2_2/(\gamma z-T)^2}}
 \left[T-\rho_2^2\frac{(\gamma z-T)}{(\gamma z-T)^2}\right]
\right\} +\ldots
\nonumber
\label{prop1}
\end{eqnarray}
Omitted terms have turned out to be of order $O(\alpha_s)$.

Let us at first consider the toy example of the  Eq.(\ref{formula1})
without extra integration over $\gamma$ which brings only technical
complexities:
\begin{equation}
 \int dz\frac{e^{iqz}}{z^{2n}} \int dT
\int\frac{d\rho_1^2}{\rho_1^2}(\rho_1^2)^{\mu_1}
\int\frac{d\rho_2^2}{\rho_2^2}(\rho_2^2)^{\mu_2}
\int dR e^{ipR-\frac{4\pi}{\alpha_s}(1-\frac{6}{\xi^2}) }
\frac{\Gamma(m_1)\Gamma(n)}{[(z-R-T)^2 +\rho_1^2]^{m_1} }
 \frac{\Gamma(m_2)}{[T^2 +\rho_2^2]^{m_2} }
\end{equation}

This integral diverges at $\rho\rightarrow \infty$. However, the
divergent part corresponding to instanton with size
$\rho\sim\Lambda_{QCD}$  contributes only to parton densities and not
to the coefficients in front of them. Moreover, these divergent parts
possess no imaginary part so we shall imply them subtracted in what
follows. (Strictly speaking we need $\mu-m$ subtractions of the type
$((x-R-T)^2+\rho^2)^{-2} - \rho^{-4}+2(x-R-T)^2\rho^{-6}+...$).
After performing the integrations one obtains
\begin{equation}
\pi^5\int dz\frac{\Gamma (n)\Gamma (-l)}{(z^2)^{n-l}}\int_0^1 du
\frac{u^{\mu_1+\mu_2}\bar{u}^{l-1}}{2^{m_1+m_2-1}
(1+u)^{\mu_1+\mu_2-1}}(\frac{6\pi\bar{u}^2}{\alpha_s(1+u)^2})^{\mu_1+\mu_2}
e^{ipxu-\frac{4\pi}{\alpha_s}(1-\frac{3\bar{u}^2}{2(1+u)^2})}
\label{coord}
\end{equation}
where we use the notation $l=\mu_1+\mu_2-m_1-m_2+4$ and $\bar u$ = $1-u$.
After continuation to Minkowski space the imaginary part of this
integral take the form:
\begin{equation}
2\pi^8\frac{B(n,-l)}{(Q^2)^{2-n+l}}\bar{x}^{n-2}
\frac{x^{\mu_1+\mu_2-2n+2l+2}(1+x)^{n-l-\mu_1-\mu_2}}{2^{m_1+m_2+2n-2l-3}}
(\frac{24\pi}{\alpha_s\xi^2})^{\mu_1+\mu_2-n+l}
e^{-\frac{4\pi}{\alpha_s}(1-\frac{6}{\xi^2})}
\label{mom}
\end{equation}
where $\bar x=1-x$ ,$B(n,l)=\frac{\Gamma(n)\Gamma(-l)}{\Gamma(n-l)}$
and $\xi$ is $2(1+x)/\bar{x}$ (see Eq.(\ref{saddle})). It is easy to see now
that the
 characteristic distances correspond to Eq.(\ref{xap}).

One may also account for for the $\rho$ dependence in the
argument of $\alpha_s$. Careful analysis \cite{bal93b} shows that
 $\frac{4\pi}{\alpha_s}$ should be changed to
$\frac{4\pi}{\alpha_s}+2b$ just as in our toy example of gluon-gluon DIS and
the
argument of $\alpha_s$ obeys the same saddle equation (\ref{saddle}).
Now, in order to find the $\gamma_{\ast}g$ amplitude Eq.(\ref{formula1})
we should take the real answer Eq.(\ref{prop1}) instead of our toy example.
After a considerable algebra (for details see \cite{bal93b})
we obtain the
following answer for the $\bar I I$ contribution
to the structure function of a real gluon:
\begin{eqnarray}
F_1^{(G)}(x,Q^2) &=&\sum_q e^2_q
\frac{1}{9\bar x^2}
\frac{ d^2\pi^{9/2}}{bS(\xi_\ast)[bS(\xi_\ast)-1]}
\left(\frac{16}{\xi_\ast^3}\right)^{n_f-3}
\nonumber\\
&&\mbox{}\times
\left(\frac{2\pi}{\alpha_s(\rho_\ast)}\right)^{19/2}
\!\!\exp\left[ -
\left(
\frac{4\pi}{\alpha_s(\rho_\ast)} +2b\right)
S(\xi_\ast)\right].
\label{answer}
\end{eqnarray}
To our accuracy, we find that the instanton-
induced contributions obey the Callan-Gross relation
$F_2(x,Q^2)= 2x F_1(x,Q^2)$.

The expression in Eq.(\ref{answer}) gives  the exponential correction to the
coefficient
function in front of the gluon distribution of the leading twist.
The exponential factor is exact to the accuracy of Eq.(\ref{action}).
The preexponential factor
is calculated to leading accuracy
in the strong coupling and up to corrections of order $O(1-x)$.
 The corresponding
contribution to the structure function of the nucleon is obtained
in a usual way, making a convolution of (\ref{answer}) with a
distribution of gluons in the proton at the scale $\rho_\ast^2$.
The instanton-antiinstanton contribution to the structure function
of a quark contains a similar contribution shown in Fig.8b.
The answer reads
\begin{eqnarray}
F_1^{(q)}(x,Q^2) &=&
\left[\sum_{q'\neq q} e^2_{q'} +\frac{1}{2}e^2_q\right]
 \frac{128}{81\bar x^3}
\frac{d^2\pi^{9/2} }{bS(\xi_\ast)[bS(\xi_\ast)-1]}
\left(\frac{16}{\xi_\ast^3}\right)^{n_f-3}
\nonumber\\
&& \mbox{}\times
\left(\frac{2\pi}{\alpha_s(\rho_\ast)}\right)^{15/2}
\exp\left[ -
\left(
\frac{4\pi}{\alpha_s(\rho_\ast)} +2b\right)
S(\xi_\ast)\right]
\label{qanswer}
\end{eqnarray}
However, in this case additional contributions exist of the
type shown in Fig.8c. They are finite (the integral over
instanton size is cut off at $\rho^2 \sim x^2/\alpha_s$),
 but the relevant instanton-antiinstanton separation
 $R$ is small, of order $\rho$.
This probably means that the structure of nonperturbative
contributions to quark distributions is more complicated.
The answer given in Eq.(\ref{qanswer}) presents the
contribution of the particular saddle point in Eq.(\ref{saddleeq}).

\subsection{Value of cross section and structure of final
state for instanton-induced particle production}
\vspace*{-0.35cm}
The instanton-induced contribution to the structure function of a
gluon in Eq.\ref{answer} is shown as a function of Bjorken $x$ for
different values of $Q\sim 10-100 GeV$ in Fig.9.
The contribution of
the box graph  is shown by dots for comparison.
The low boundary for possible values of $Q$ is determined by
the condition that the effective instanton size is not too
large. At $Q=10 \,GeV$ we find $\rho_\ast \simeq 1\, GeV^{-1}$, cf.
Fig.7. This value is sufficiently small, so that instantons
are not distorted too strongly by large-scale vacuum fluctuations.
Another limitation is that the valley approach to the
calculation of the "holy grail" function $S(\xi)$ is likely to be
justified at $S(\xi) \ge 1/2 $, which
translates to the condition that $x>0.3-0.35$.
\begin{figure}[t]
    \begin{center}
\begin{picture}(150,350)
\end{picture}
    \end{center}
\caption[xxx]{  \small
                  Nonperturbative contribution to the structure function
                  $F_1(x,Q^2)$ of a real gluon (\ref{answer})
                  as a function of $x$ for
                   different values of $Q$ (solid curves).
                   The leading perturbative contribution
                   is shown for comparison by dots. The dashed curves
                   show lines with the constant effective value of
                   the action on the $\bar I I$ configuration.
 }
\end{figure} \nopagebreak
Numerical results are strongly sensitive to the particular value
of the QCD scale parameter.
 We use the two-loop expression for the coupling
with three active flavors,
and the value $\Lambda_{\overline MS}^{(3)} = 365 MeV$ which
corresponds to the coupling at the scale of $\tau$-lepton mass
$\alpha_s (m_\tau)= 0.33$ \cite{ALEPH}.
Since the dependence on the coupling is exponential, the 20\% increase
of $\alpha_s(\rho_\ast)$ induces the increase of the cross section
by almost an order of magnitude! Together with uncertainties in the function
$S(\xi)$ and in the preexponential factor, this indicates that the
particular curves given in Fig.9 should not be taken too seriously,
and rather give a target for further theoretical (and experimental?)
studies to shoot at.

To summarize, we have found that
 instantons produce a well-defined and calculable
contribution to the cross section of deep inelastic scattering
 for sufficiently large values of $x$ and large $Q^2\sim 100 - 1000
GeV^2$,
which turns out, however, to be rather small ---
of order $10^{-2}-10^{-5}$ compared to the
perturbative cross section.
This  means that the
accuracy of standard perturbative analysis is sufficiently
high, and that there is not much hope to observe
the instanton-induced contributions to the total deep inelastic
cross section experimentally.
However, instantons are likely to produce events with
a very specific structure of the final state, and such
peculiarities may be subject to experimental search.
The dominating Feynman diagrams in our calculation correspond to
 $2\pi/\alpha(\rho_\ast)\sim 15 $ gluons and $2n_f-1 =5$ quarks
in the final state with the low energy of order $\rho_\ast^{-1}
\sim 1\,GeV$. They  are produced in the spherically symmetric way
 in the c.m. frame of the partons colliding
through the instanton.( The transverse momentum of the quark
coming to the instanton is $k_{\perp}^2\sim Q^2\alpha_s \sim $ few Gev
 and so is the transverse momentum of the current quark jet).
It is not likely that quarks and gluons emitted from the instanton
can be resolved as minijets (they have  $k_{\perp}\sim Q\alpha_s \sim
$ 1 Gev), and we rather expect a spherically symmetric production of
final state hadrons in this frame. The effect is likely to be
resonance-like, that is present in  a narrow interval of values
of Bjorken $x$ of order 0.25--0.35 in the parton-parton collision.
\newpage
\subsection{Instanton-induced particle production in other hard processes}
\vspace*{-0.35cm}
  Here we consider another hard process - hadron-hadron scattering with large
$p_{\perp}$ in the final state. Although the complete analysis of the
instanton-induced particle production for this case is far from finished, first
results indicate that these instanton -induced contributions are indeed
well-defined and the size of instantons is also of order of $1/Q\alpha_s(Q)$
where Q is a scale of transverse momentum.

 We consider the inclusive cross section of a gluon-gluon scattering at high
energies with production of a single gluon jet in the final state (see Fig.10).
For simplicity we consider the case of $90^o$-scattering,and take $p_1=(p,\vec
p)$, $p_2=(p,-\vec p)$, $q=(q,\vec q)$ and $\vec{p}\cdot\vec{q}=0$. The
kinematics is that $s=2p^2$ and $\vec{q}^2$ are large ($\gg\Lambda_{QCD}$).
\begin{figure}
    \begin{center}
        \begin{picture}(100,150)
        \end{picture}
    \end{center}
    \caption[xxx]{ \small
                   The instanton-induced contribution to the
                   gluon-gluon collision with a production of a
high-$p_{\perp}$ jet                    }
\end{figure}

The idea is that the hard-hard and hard-soft quantum corrections shown in
Fig.10 will
provide us with a cutoff for the $\rho$ integrals. The hard-hard correections
are given by the formula \cite{MU91}
\begin{equation}
exp{(-{\alpha_s\over 4\pi}\rho^2\sum_{i<j} p_i\cdot p_j \ln{\Theta_{ij}})}
\label{hard}
\end{equation}
where the summation goes over all the hard particles and $\Theta_{ij}$ is the
angle between their momenta.
Since we do not
know how to calculate the soft-hard corrections we have adopted a
model for them, namely that they coincide with the hard-hard ones for
the case where each soft gluon carries the momentum $k/n$ where $k=p_1+p_2-q$.
Then
\begin{eqnarray}
&-\sum_i(p_1k_i) \ln{(p_1k_i)}~-~\sum_i (p_2k_i)
\ln{(p_2k_i)}~+~\sum_i (qk_i) \ln{(qk_i)}~+~\sum_{i<j}(k_ik_j)
\ln{(k_ik_j)}&\nonumber\\
&=~-\sum_i(p_1{k\over n}) \ln{(p_1{k\over n})}~-~\sum_i (p_2{k\over n})
\ln{(p_2{k\over n})}~+~\sum_i (q{k\over n}) \ln{(q{k\over
n})}~+~\sum_{i<j}({k\over n})^2 \ln{({k\over n})^2}&\nonumber\\
&=~-~(p_1k)\ln{(p_1k)}~-~(p_2k)\ln{(p_2k)}~+~(qk)\ln{(qk)}~+~{k^2\over
2}\ln{(k^2)}&
\label{soft}
\end{eqnarray}
so at least $n$ disappears from the final answer which we think signals that
the model is not unreasonable.
  With this model for soft-hard corrections we have (cf. Eq(\ref{1q})):
\begin{equation}
\int \rho\,d\rho\int dR_0\,\exp\left\{-{\alpha_s\over
2\pi}M^2(\rho_1^2+\rho_2^2)+iER_0-
\frac{4\pi}{\alpha_s(\rho)}S(\xi)\right\}.
\label{j1}
\end{equation}
where $E$ is now $\sqrt{k^2}=2\sqrt{p(p-q)}$. Here $M^2$ is the sum of
hard-hard corrections (calculated using the Eq.(\ref{hard}) for the three hard
particles) and of the soft-hard corrections Eq.(\ref{soft}):
\begin{equation}
M^2~=~s\ln{2} -{1\over 2}s\ln{(1+{E^2\over s})}+{E^2\over 2}\ln{2}-{E^2\over
2}\ln{(1+{s\over E^2})}
\end{equation}
At $E^2=0$ $M^2=s\ln 2$ and it monotonously decreases to $M^2={1\over 2}s\ln 2$
while $E^2$ increases towards $E^2=s$. It is easy to see now that
the integrals over $\rho$ converge at $\rho\sim (E\alpha_s(E))^{-1}$ where
$E=\sqrt{(p_1+p_2)^2}$. Similarly to the case of deep inelastic scattering (see
Eq.(\ref{1q})) the integral over $R_0$ can be removed into the complex plane to
a `Euclidean' saddle point given by equations:
\begin{eqnarray}
 E\rho_\ast &=&
\frac{8\pi}{\alpha_s(\rho_\ast)}\sqrt{\xi_\ast-2}S'(\xi_\ast) ,
        \nonumber\\
M^2\rho_\ast^2\left( {\alpha_s\over \pi}+{b\over 4}({\alpha_s\over
\pi})^2\right)
  &=&\frac{8\pi}{\alpha_s( \rho_\ast)}
(\xi_\ast-2)\,S'(\xi_\ast)+2bS(\xi_\ast)\,,
\label{saddleq}
\end{eqnarray}
where $\rho$ is any of the $\rho_1, \rho_2$. It is clear now that the saddle
value of $\rho$ corresponds to the general rule Eq.(\ref{size}). The analog of
$\epsilon$ for this problem is
$E^2/s$ so as in the case of deep inelastic scattering we can start from the
region where soft gluons carry a small fraction of total energy and move using
the valley formulas up to "Zakharov's point" where we cancel half of the
initial Gamow factor. We believe that for this process the ratio
signal/background for the instanton-induced contribution will be better than
for deep inelastic scattering. The only thing left is to calculate actually
these soft-hard corrections (at least in the region of small $E^2/s$). The
study is in progress.

\section{Conclusions}

I think that for any hard process in QCD there is a non-perturbative
contribution coming from the small-size instantons with
\begin{equation}
\rho~\sim ~{1\over Q\alpha_s(Q)}
\label{size}
\end{equation}
where $Q$ is a characteristic scale of a hard process. For the two specific
hard processes studied above the numerical difference between the
characteristic scale $Q$ and the instanton size $\rho^{-1}$ is even greater:
typically there is a factor $\sim 3$ in the numerator in r.h.s. of
Eq.(\ref{size}). These non-perturbative cross sections are closely related to
the instanton-antiinstanton interactions (this relation shown in Fig.1 can be
formalized by the introducing the notion of instanton-induced effective
Lagrangian, see refs.\cite{cdg,svz,NTTL}). These \crse s become of interest
only at relatively large energies which correspond to the interaction of
strongly overlaped $I$ and $\bar{I}$, and we have no quantitative description
of the instanton interactions in this situation. Therefore in our papers
(including this one) we  rather set the problem of instanton-induced cross
sections in QCD than solve it.
 From our estimates we can only deduce that the instanton-induced
 \crse s appear to be a steep functions of the ratio of energy which
has `flowed into the instanton' to the characteristic scale of a hard process
and when these functions became (relatively) large we lose the theoretical
control over our calculations. However, we believe that finding an
instanton-induced particle production
at high energies is a challenging problem, and further theoretical
efforts are needed to put it as a practical proposal to
experimentalists.

\section{Acknowledgements}
I would like to thank V.M.Braun for a very rewarding collaboration.
It is a pleasure to
acknowledge also useful discussions with
J.C.Collins, D.I.Diakonov, E.M.Levin, A.H.Mueller, V.Yu.Petrov and M.S.Ryskin.
This work was supported by the US Department of Energy
under the grant DE-FG02-90ER-40577.

\section{References}


\vglue 0.5cm

\end{document}